\documentstyle[12pt]{article}
\begin{document}
\def\Tm{T^m}
\def\tm{t_m}
\def\Tdm{T^m_d}
\def\tdm{t_m^d}
\def\ot{\otimes}
\def\br{\bf R}
\def\al{\alpha}
\def\de{\delta}
\def\De{\Delta}
\def\lm{\lambda}
\def\b{\beta}
\def\dg{\dagger}
\def\k{\kappa}
\def\om{\omega}
\def\si{\sigma}
\def\w{\wedge}
\def\Tn{T^n}
\def\m{\sl m}
\def\tn{t_n}
\def\Tdn{T^n_d}
\def\tdn{t_n^d}
\def\e{\varepsilon}
\def\ti{\tilde}
\def\I{{\cal I}}
\def\H{{\cal H}}
\def\C{{\cal C}}
\def\G{{\cal G}}
\def\G{{\cal G}}
\def\d{\partial}
\def\ug{U(\G)}
\def\uqg{U_q(\G)}
\def\la{\langle}
\def\ra{\rangle}
\def\lta{\leftarrow}
\def\rta{\rightarrow}
\def\ot{\otimes}

\def\be{\begin{equation}}
\def\ee{\end{equation}}
\def\jp{{1\over 2}}
\def\De{\Delta}
\def\tD{\Delta^*}

\begin{flushright}
{}~
IML 03-06\\
hep-th/0306114
\end{flushright}

\vspace{1pc}
\begin{center}
{\Huge \bf On Hopf anomaly}\\
[25pt]{\small
{\bf Ctirad Klim\v{c}\'{\i}k}
\\ ~~\\Institute de math\'ematiques de Luminy,
 \\163, Avenue de Luminy, 13288 Marseille, France}
\end{center}

\vspace{0.2 cm}
\centerline{\bf Abstract}
\vspace{0.2 cm}
 This is  a very brief  but selfcontained review of
the concept of
 quantum group symmetries and  their anomalies.
  Remarkably, general constructions can be very
simply illustrated
on the standard harmonic oscillator which is shown to possess
a  non-commutative and non-cocommutative anomalous quantum
group symmetry.

\section{Introduction.}
The goal of these notes is to illustrate  some Hopf
(or quantum group)
symmetry   story on a simple example. For the latter,
we have chosen
 the standard  harmonic oscillator. General quantum group
facts presented here
 are  mostly known (though they are often presented
in the literature in a somewhat scattered way) but the
harmonic oscillator example  of
  Hopf anomaly is original.

In the second section, we define
 the notion
of the Hopf  symmetry of quantum dynamical systems and in the
third section, we develop the concept of Hopf
 moment maps and Hopf anomalies.
In the fourth  section, we illustrate all those notions on the
simple example: the scaling symmetry of the  harmonic oscillator.
 Some definitions and basic properties of
  Hopf algebras can be found
  in Appendix.
\section{Hopf  symmetry}
Recall that an {\em abstract}  quantum dynamical system is a pair
$(A,\al_t)$ consisting of a noncommutative algebra $A$
of observables and a time-evolution, i.e. a one-parameter
 group $\al_t$ of
automorphisms of $A$.   We say that a
 Hopf algebra $\H$ is the Hopf
symmetry of $(A,\al_t)$ if:
\vskip1pc
\noindent 1. $A$ is a $\H$-module-algebra.

\noindent 2. Given $x\in\H$, it exists $x(t)\in \H$ such that
there is the following equality of maps $A\to A$:
$\al_t\circ  x =x(t)\circ \al_t$.

\vskip1pc
\noindent {\bf Clarifications}: 1. The algebra $A$ is the
$\H$-module-algebra if $\H$ acts on $A$ (i.e. there is
the morphism of {\em algebras} $\H$ and $End(A)$) and this action
respects  the following {\it Hopf-Leibniz} rule:
$$x(f*g)= (x'f)*(x''g),\quad x\in \H, \quad f,g\in A.\eqno(1)$$
Here we used the Sweedler notation for the
coproduct $\De x=x'\ot x''$ and $*$ denotes the
noncommutative product in $A$.
The Hopf-Leibniz rule can be interpreted as follows:
the coproduct on $\H$ encodes
 the compatibility of the $\H$-action with the product
$*$ in $A$.
From the point of view
of noncommutative geometry, the Hopf symmetry thus acts
by "noncommutative vector fields" which are NOT necessarily
the derivations of the noncommutative algebra $A$. This
may seem surprising but it is the fact which has been since
long time
established either in  the noncommutative
geometry literature by [Connes and Moscovici 1] or
on the quantum group side, see  e.g. the book of
[Majid].    Actually, only when we take   $\H$ to be
the (cocommutative)  envelopping algebra $\ug$
of some Lie algebra $\G$,  we can see from (A1) that
the Hopf-Leibniz rule (1) implies the action of
  $\G$  on $A$ by derivations. Indeed, Eqs.(1)  and (A1) then give
$$x(f*g)=(xf)*g +f*(xg),\quad x\in\G,\quad f,g\in A.$$
In fact,  this special
 case $\H=\ug$ corresponds to the situation when the Hopf symmetry
becomes the ordinary symmetry known from the textbooks on
quantum mechanics.

\vskip1pc
~~~~~~~~~~~~~~~~~~~ 2. The second condition of the Hopf symmetry
is sometimes formulated
in a more restrictive  way by claiming that
  the action of $\H$ on $A$
   should commute with the evolution   $\al_t$.
  However, we prefer  our more general
 formulation which, by the way, is necessary
     to   describe
  important symmetries as e.g. the Poincare   symmetry of
relativistic systems.

\section{Hopf moment maps and anomalies}
The Hopf symmetry $\H$ of $(A,\al_t)$ is called
Noetherian if it admits the Hopf moment map. If it does not
admit it, it is called anomalous. To our best knowledge,
 the concept of the
   Hopf moment
map was introduced by [Korogodsky].
 It is  an {\em algebra} homomorphism   $m:
\H\to A$ such that  the action of
$\H$ on $A$ can be written as follows
$$\label{Presov}xf=m(x')*f*m(S(x'')), \quad x\in \H,
 \quad f\in A.\eqno(2)$$
Here again we use the Sweedler notation for the
coproduct and $S$ stands for
the antipode of $\H$.
In particular, if $\H=\ug$ and
 $x\in\G$, then the formulae for the coproduct
and the antipode (A1)  permit to rewrite (2) in
the following familiar
form
$$xf= m(x)*f-f*m(x)\equiv [m(x),f].$$
The Noetherian Hopf symmetry descends to the
 space of states $V$. Let us explain what this means:
we   call a triple $(A,\al_t,V)$ a {\em concrete} realization
of the  {\em abstract} quantum
dynamical system $(A,\al_t)$ if $V$ is the representation space
of the algebra
$A$ such that the evolution
 is given by unitary operators
$U(t)$ on $V$. The  Hopf  moment map $m$  gives
immediately the representation of $\H$ on $V$.
On the other hand,
 the anomalous Hopf
symmetry exists only on the level of
the algebra of observables and cannot be implemented on
 the space of states.

\section{Scaling the harmonic oscillator}

Consider the  anihilation and creation operators $a,\bar a$
fulfilling the standard commutation relation
$$[a,\bar a]=1.$$
The algebra $A$   consists of polynomials in $a,\bar a$
and the evolution automorphism $\al_t$ on $A$ is  defined by
$$\al_t(a)=e^{-i\om t}a,\quad \al_t(\bar a)=e^{i\om t}\bar a,$$
where $\om$ is a parameter.

Then consider a Hopf algebra $H$ (with a unit element $\eta$)
 generated by   $T_0,T_1,T_2$ and by the following
relations
$$ [T_0,T_1]=2T_1,\quad [T_0,T_2]=2T_2,\quad [T_1,T_2]=0.$$
The coproduct $\Delta$ is defined as
$$ \Delta T_\alpha=\eta\ot T_\al +T_\al\ot \eta,\quad
\al=1,2;\eqno(3)$$
$$ \Delta T_0=\eta\ot T_0 +T_0\ot \eta-4 T_1\ot T_2.\eqno(4)$$
The antipode $S$ reads
$$ S(\eta)=\eta,\quad S(T_0)=-T_0-4T_1T_2,\quad S(T_1)=-T_1,
\quad S(T_2)=-T_2\eqno(5)$$
and the counit $\e$ is
$$ \e(\eta)=1, \quad \e(T_0)=\e(T_1)=\e(T_2)=0.$$
{\bf Note}: Although  our Hopf algebra $\H$ does not coincide
with that of [Connes and Moscovici 2] there are nevertheless
some similarities between them.
\vskip1pc
\noindent Now we define the action of $\H$ on $A$ as follows
$$T_0(\bar a^m a^n)=-2(m+n)\bar a^m a^n,\quad T_1(\bar a^m a^n)=
-n\bar a^m a^{n-1},\quad T_2(\bar a^m a^n)=m\bar a^{m-1} a^n.$$
Is this $\H$-action the Hopf symmetry of $(A,\al_t)$? It is easy to
  check   the Hopf-Leibniz rule (1) with the help of the
following formulae:
$$T_0f=  -2[\bar aa,f]+4[\bar a,f]a,\quad T_1 f=[\bar a,f],\quad T_2
f =[a,f] ,
\quad f\in A.\eqno(6)$$
 It is also
easy to verify the second condition on p.2. Indeed, the action
of $T_0$ clearly commutes with the evolution $\al_t$, while for
the action of $T_1$ and  $T_2$, we have
$$\al_t\circ T_1=(e^{i\om t}T_1)\circ \al_t,\quad
 \al_t\circ T_2=(e^{-i\om t}T_2)\circ \al_t.$$
Thus $\H$  is  the Hopf symmetry of the harmonic
oscillator.
\vskip1pc
\noindent It turns out that the Hopf symmetry $\H$ does not
admit the Hopf moment
map $m$. Indeed, by inspecting the formulae (2), (3), (4), (5) and (6),
we easily find that
if $m$ existed then it would have to  be given by
$$m(T_0)=-2\bar aa+c_0,\quad m(T_1)=\bar a+c_1,\quad m(T_2)=a.\eqno(7)$$
Here $c_0,c_1$ are central in $A$.
 However, whatever choice of   $c_0,c_1\in A$ we make,
 the map $m$ defined by Eq.(7) is not the algebra homomorphism
because
$$m([T_1,T_2])\neq [m(T_1),m(T_2)],\quad m([T_0,T_1])\neq
[m(T_0),m(T_1)].$$
   What is the physical interpretation of this Hopf symmetry?
Clearly, $(T_1+T_2)$ and $i(T_1-T_2)$ generate
 the Galileo boost and  the translation, respectively,
while $T_0$ is just the scaling
of $A$.  We know that the classical Hamiltonian equations
of the harmonic oscillator are linear; this is also true
for the quantum equations of motion written in the Heisenberg
picture:
$${d\over dt}a(t)=-i\om a(t),\quad {d\over dt}\bar a(t)=i\om \bar a(t).$$
It is therefore obvious, that the   scaling  $a,\bar a\to \lm a,\lm\bar a$
commutes with the evolution
$\al_t(a)\equiv a(t),\al_t(\bar a)\equiv \bar a(t)$.
 We have moreover established  that this scaling
is the anomalous non-cocommutative Hopf symmetry.

\section{Appendix: A Hopf primer}
\noindent {\bf Definition}: A Hopf algebra $\H$ is
an unital associative algebra equipped with algebra homomorphisms
$\Delta: \H\to \H\ot \H$ and $\e:\H\to\br$
and with an algebra antihomomorphism $S:\H\to\H$ (i.e $S(xy)=S(y)S(x)$,
$S(\eta)
=\eta$)
verifying
the following axioms
$$\e(x')x''=x'\e(x'')=x,\quad S(x')x''=x'S(x'')=\e(x)\eta,
\quad \e(S(x))=\e(x),$$
$$\quad S(x')\ot S(x'')=
(S(x))''\ot (S(x))',\quad
\Delta x'\ot x''= x'\ot \Delta x'',\quad x\in\H.$$
\vskip1pc
\noindent {\bf Clarifications}:

\noindent 1. We have denoted the   unit of $\H$ by $\eta$.

\noindent 2. The multiplication in $\H$ can be  denoted   as
$\mu:\H\ot\H\to\H$ but often we write only
$xy$ instead of $\mu(x\ot y)$.

\noindent 3. We have used the Sweedler notation for the coproduct
$$\Delta x=\sum_j x'_j \ot x''_j\equiv x'\ot x'', \quad
x\in \H.$$

\noindent 4. We note that the product on $\H$    canonically
induces the product on $\H\ot \H$ whose unit
element is $\eta\ot \eta$.

\noindent 5. We say that $\H$ is cocommutative if
$x'\ot x''=x''\ot x'$.

\vskip1pc
\noindent {\bf Basic cocommutative example}:
  The envelopping algebra $U(\G)$
of a Lie algebra $\G$ is the cocommutative Hopf algebra whose structure
is completely determined by the Lie bracket $[.,.]$ in $\G$. Indeed, if $T$
is the tensor algebra of the vector space $\G$, then $U(\G)=T/J$,
where the two-sided ideal $J$ is generated by  tensors
$$x\ot y-y\ot x -[x,y], \quad x,y\in\G.$$
The coproduct $\Delta: U(G)\to U(\G)\ot U(\G)$, the counit $\e:\ug\to\br$
and the antipode $S:\ug\to\ug$ are defined as
 $$   \Delta x=\eta\ot x+x\ot \eta,\quad  \e(x)=0,\quad S(x)=-x,\quad
x\in \G . \eqno(A1)$$
We note that the definition of $\Delta$, $\e$ and $S$ acting on whatever
element of $U(\G)$ follows from (A1) and from the
(anti)homomorphism properties of $\Delta$, $\e$ and $S$.

\vskip2pc
\noindent {\large  \bf Acknowledgement}:
\noindent  I thank to my colleagues from the Luminy seminar for having
encouraged
me to write down these notes.
\vskip2pc
\noindent {\large  \bf References}:

\noindent [Connes and Moscovici 1] A.Connes and P. Moscovici, {\it Cyclic
cohomology

\noindent ~~~~~~~~~~~~~~~~~~~~~~~~~~~~~~~~and Hopf symmetry},
{\it Lett.Math.Phys.} {\bf 52} (2000) 1,

\noindent ~~~~~~~~~~~~~~~~~~~~~~~~~~~~~~~~~~preprint math.OA/0002125

\noindent [Connes and Moscovici 2] A.Connes and P. Moscovici, {\it Hopf
algebras, cyclic }

\noindent ~~~~~~~~~~~~~~~~~~~~~~~~~~~~~~~~~~{\it cohomology and transverse
index theorem},

~~~~~~~~~~~~~~~~~~~~~~~~~~~~~{\it Commun. Math. Phys.} {\bf 198} (1998) 199,

~~~~~~~~~~~~~~~~~~~~~~~~~~~~~ preprint math.DG/9806109

\noindent [Korogodsky]~~~~~~~~~~~~~~~~~~~L. Korogodsky: {\it Complementary
series representa-

~~~~~~~~~~~~~~~~~~~~~~~~~~~~tions and quantum
orbit method},

~~~~~~~~~~~~~~~~~~~~~~~~~~~~~ preprint q-alg/9708026
~

\noindent [Majid] ~~~~~~~~~~~~~~~~~~~~~~~~S. Majid: {\it Foundations
of quantum group theory}~,

~~~~~~~~~~~~~~~~~~~~~~~~~~~~~~Cambridge University Press 1995

\end{document}